\documentclass[superscriptaddress,nobinotes,twocolumn,amsmath,amssymb,prl]{revtex4-1}

\usepackage{pdfpages} 
\usepackage{etoolbox} 

\usepackage{bm,mathptmx,braket}

\usepackage{graphicx,color,hyperref}
\usepackage[caption=false]{subfig}
\hypersetup{colorlinks=true, linkcolor=blue, citecolor=blue, urlcolor=blue} 

\newcommand{\beq}[1]{\begin{equation}\label{#1}}
\newcommand{\eep}{\;.\end{equation}}
\newcommand{\eec}{\;,\end{equation}}
\newcommand{\eeq}{\end{equation}}

\newcommand*\dd{\mathop{}\!\mathrm{d}} 

\newcommand*{\heading}[1]{\belowpdfbookmark{#1}{#1}{\textit{#1.---}}\ignorespaces}
\let\section\heading 

\newcommand*\chem[1]{\ensuremath{\mathrm{#1}}} 

\renewcommand{\b}{\beta}

\newcommand{\ep}{\epsilon}

\newcommand{\s}{\sigma}
\newcommand{\p}{\phi}

\DeclareMathAlphabet{\mathcal}{OMS}{cmsy}{m}{n} 

\newcommand{\F}{\mathcal{F}}    

\newcommand{\dc}{d_{\text{c}}}

\makeatletter
\patchcmd{\@outputpage@head}{\@ifx{\LS@rot\@undefined}{}{\LS@rot}}{}{}{}
\makeatother

\begin{document}


\title{
Pseudo-proper two-dimensional electron gas formation
}


\newcommand{\TCM}{{Theory of Condensed Matter, Cavendish Laboratory, University of Cambridge, J.\,J.\,Thomson Avenue, Cambridge CB3 0HE, UK}}
\newcommand{\HarvardSeas}{John A.~Paulson School of Engineering and Applied Sciences, Harvard University, Cambridge, Massachusetts 02138, USA}
\newcommand{\nanoGUNE}{CIC Nanogune and DIPC, Tolosa Hiribidea 76, 20018 San Sebastian, Spain}
\newcommand{\Ikerbasque}{Ikerbasque, Basque Foundation for Science, 48011 Bilbao, Spain}
\newcommand{\Durham}{Centre for Materials Physics, Durham University, South Road, Durham DH1 3LE, UK}


\author{Daniel Bennett}
\email{dbennett@seas.harvard.edu}
\affiliation{\TCM}
\affiliation{\HarvardSeas}
 
\author{Pablo Aguado-Puente}
\affiliation{\nanoGUNE}

\author{Emilio Artacho}
\affiliation{\TCM}
\affiliation{\nanoGUNE}
\affiliation{\Ikerbasque}
 
\author{Nicholas C.~Bristowe}
\affiliation{\Durham}

\date{\today}

\begin{abstract}
In spite of the interest in the two-dimensional electron gases (2DEGs) experimentally found at surfaces and interfaces, important uncertainties remain about the observed insulator--metal transitions (IMTs).
Here we show how an explicit pseudo-proper coupling of carrier sources with a relevant soft mode significantly affects the transition.
The analysis presented here for 2DEGs at polar interfaces is based on group theory, Landau-Ginzburg theory, and illustrated with first-principles calculations for the prototypical case of the \chem{LaAlO_3/SrTiO_3} interface, for which such a structural transition has recently been observed.
This direct coupling implies that the appearance of the soft mode is always accompanied by carriers. For sufficiently strong coupling an avalanche-like first-order IMT is predicted.
\end{abstract}

\maketitle


\section{Introduction}
\begin{figure}[t!]
\centering
\includegraphics[width=\columnwidth]{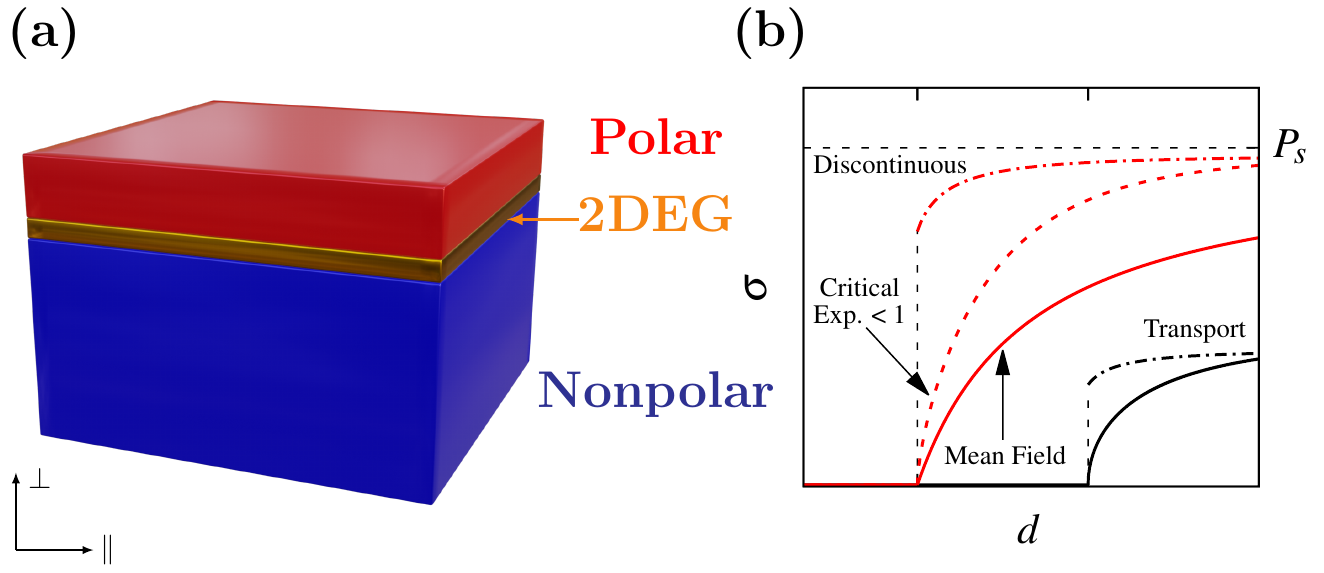}
\caption{
{\bf (a)} Sketch of 2DEG formation between a polar film and a nonpolar substrate.
{\bf (b)} Illustration of the possibilities for the onset of interfacial carriers with film thickness. The red lines show the 2DEG carrier concentration, for a mean field theory (solid), beyond mean-field (dashed), and a discontinuous transition (dot-dashed). The black lines represent the onset of conductivity, which may differ from the onset of carriers.
}
\label{Fig1}
\end{figure}

The insulator--metal transition (IMT) is one of the most widely studied problems in condensed matter physics \cite{mott1968metal,imada1998metal}, particularly in oxide materials \cite{morin1959oxides,adler1968mechanisms,yang2011oxide}. 
IMTs in two dimensions (2D) have attracted additional interest \cite{kravchenko1995scaling,kravchenko2003metal}, in particular the one occurring at polar-nonpolar perovskite interfaces, first observed for thin films of \chem{LaAlO_3} (LAO) grown on \chem{SrTiO_3} (STO) \cite{ohtomo2002,ohtomo2004,chenlaalo3}. 
For film thickness beyond a critical value (3 unit cells), a 2D electron gas (2DEG) appears at the interface, or can be induced by gating \cite{bi2016electro,yang2020nanoscale}, see Fig.~\ref{Fig1} (a). The gas is associated to Ti 3$d$ states, and has been found to display intriguing magnetic \cite{2deg_magnetism} and superconducting \cite{2deg_superconductivity} properties, even in co-existence \cite{2deg_coexistence_1,2deg_coexistence_2}, in addition to non-trivial topological character \cite{vistoli2019giant}.
2DEG formation has also been proposed as an alternative screening mechanism to domains \cite{kmf,bennett2020electrostatics} in ferroelectric materials \cite{pablo_2deg_theory,pablo_2deg_dft}, supported by recent experimental evidence \cite{2deg_2018,2deg_2018_2,brehin2023coexistence}.
In addition, similar observations of surface carriers have been made in stacks of 2D materials \cite{deb2022cumulative,cao2024polarization},
which have recently been predicted \cite{li2017binary,bennett2022electrically,bennett2022theory,bennett2023polar,bennett2023theory} and experimentally shown \cite{yasuda2021stacking,yasuda2024ultrafast} to be ferroelectric.

If both insulators are centrosymmetric, as is the case in LAO/STO, the polar discontinuity is quantized \cite{stengel2009berry,bristowe_electronic_reconstruction} in units of $P_S = P_0/2$, i.e.~half a quantum of polarization, or 0.5 electrons per interface Ti.
Although electronic reconstruction -- electrons from the valence band at the surface becoming more stable at the bottom of the conduction band at the interface -- was initially proposed as the source of carriers, redox defects were soon recognized as a more likely source \cite{cen2008nanoscale,zhong2010polarity,superlattice_defect,yang2023coexistence}, most prominently oxygen vacancies.
This has been amply ratified by various alternative experimental methods to generate very similar 2DEGs, namely, by irradiation \cite{santander2011two} and chemistry (controlling the oxygen partial pressure \cite{2deg_strain}, or depositing oxygen-hungry metallic layers on STO \cite{rodel2016universal}), in addition to detailed first-principles calculations \cite{zunger_dft_paper}.

Despite considerable interest, some aspects of 2DEG formation and the IMT at polar--nonpolar interfaces are still unclear. 
For example, it is not clear whether the appearance of carriers with film thickness or electric field is continuous, as predicted in a mean-field approximation \cite{bristowe_electronic_reconstruction}, or discontinuous, as
either assumed \cite{zunger_dft_paper}, or postulated beyond mean-field \cite{superlattice_defect}, see Fig.~\ref{Fig1} (b). 
Additionally, the distinction between the onset of carriers and the onset of conduction is not typically made. 
It may be possible that carriers can appear but are not conducting. 
This is supported by experimental evidence: a density of trapped Ti 3d-like states has been observed in LAO/STO at just 2 unit cells of LAO, i.e.~before the observation of the mobile carriers \cite{sing2009profiling}. 
For carriers generated via oxygen vacancies, the charge carriers are trapped below the surface defects which generated them, and thus will be localized in the plane of the interface \cite{superlattice_defect}. 
As the thickness of the film increases, more surface defects will be created and the associated carriers will become delocalized, and conductivity will occur at a thickness larger than the one at which carriers first appear, see Fig.~\ref{Fig1} (b). 
It has also been proposed that both heavy and light carriers form, with only the lighter carriers contributing to conduction \cite{popovic2008origin}.

\begin{figure}[t!]
\centering
\includegraphics[width=\columnwidth]{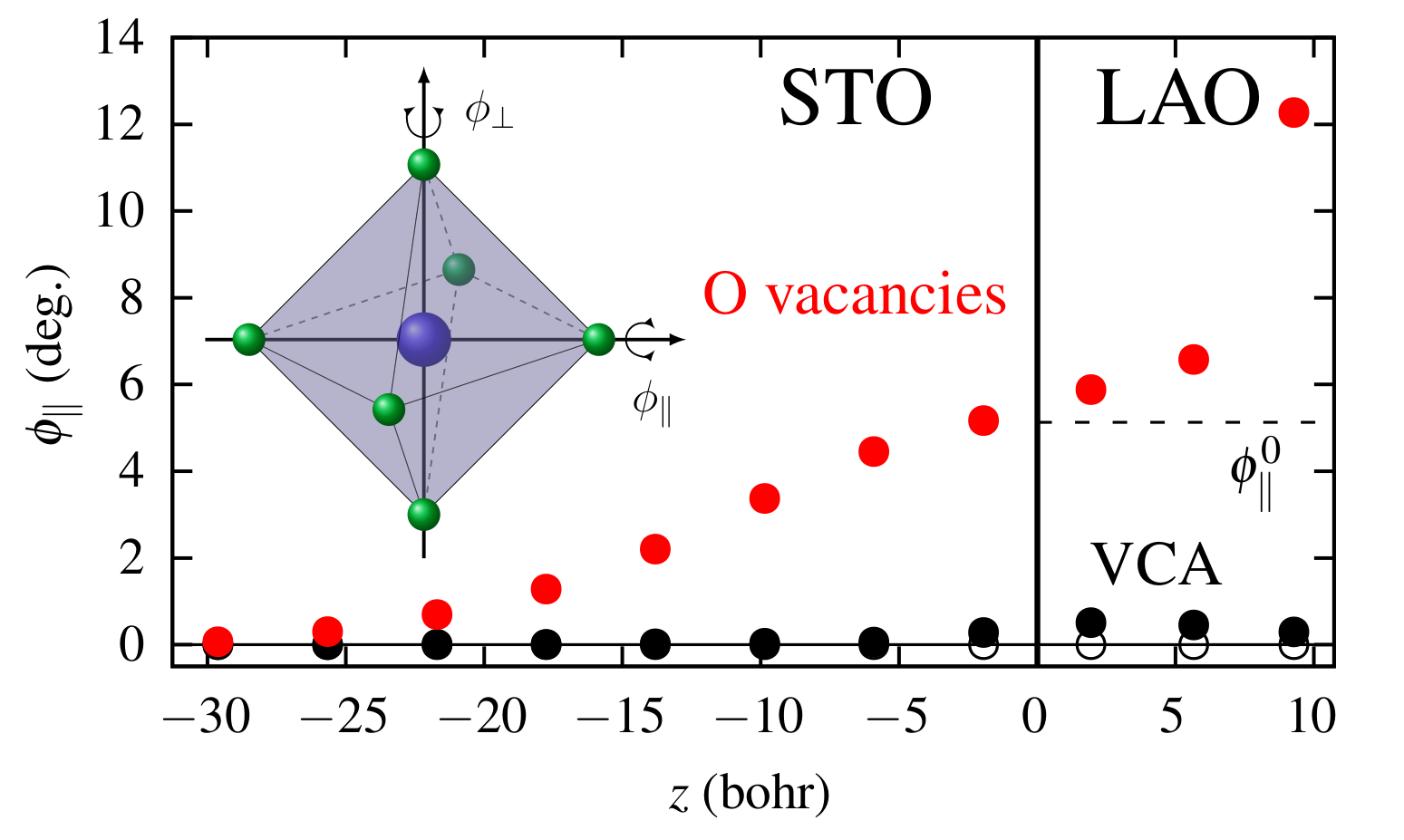}
\caption{
$\p_\parallel$ tilt profile in LAO/STO from first-principles calculations. The hollow points show results with no 2DEG (${\s'=0}$), the filled points show results with a fully saturated 2DEG (${\s' = 1}$) using VCA (black) and explicit oxygen vacancies (red).
The horizontal line on the LAO side of the interface indicates the magnitude to the tilts in bulk strained LAO, $\p^0_\parallel$.
The inset shows a sketch of an \chem{AlO_6} octahedron. The axes of the $\p_{\parallel}$ and $\p_{\perp}$ tilts, [100] and [001] respectively, are shown.
}
\label{Fig2}
\end{figure}

It was recently proposed that the coupling between the dielectric properties and structural instabilities, such as octahedral rotations (tilts, see inset in Fig.~\ref{Fig2}), may play a role in the onset of carriers, and possibly the IMT, and even change the character of the transition, or lead to a sequence of transitions \cite{bennett-LAO}. 
Tilting has been experimentally shown to occur in LAO/STO, where a transition with thickness was recently observed \cite{stengel_tilt,song2021electronic,mun2023extended}.

Such instabilities may play a deeper role in the IMT when considering oxygen vacancies as the source of carriers. In addition to providing free charge which can migrate to the interface, removing an oxygen atom will break an octahedron at the surface, which may strongly affect the tilting and polar modes.
In order to better understand the rich coupling between such soft-mode instabilities, defects and interfacial carriers, we perform large-scale first-principles calculations of the LAO/STO interface. From a group theory analysis, we find that there is an additional direct coupling between octahedral rotations and oxygen vacancies at the surface, which propagates into the film and substrate. Using a Ginzburg-Landau (GL) theory, we show that this direct coupling at the surface can strongly affect the onset of interfacial carriers, and can drastically change the order of the IMT.

\section{First-principles calculations} 
Calculations were performed using the {\sc siesta} code \cite{siesta} on vacuum-terminated \chem{(SrTiO_3)_n/(LaAlO_3)_m} slabs, 
where $n=8$ or 15 unit cells, and $m = 3$ unit cells \cite{SM}. 
The vacancies were explicitly introduced by removing one of the oxygen atoms from the \chem{AlO_2} surface; 
removing one oxygen atom from a $2 \times 2$ in-plane supercell is equivalent to imposing a displacement field of $D_{\perp}=0.5$ electrons per unit cell surface area. 
In order to distinguish between the effect an oxygen vacancy has on just the electrostatic boundary conditions,
compared to the effect of symmetry breaking on structural instabilities, 
we redo the calculations using the virtual crystal approximation (VCA),
substituting all of the oxygen atoms on the surface with `virtual atoms' with fractional atomic numbers to
impose the same electric displacement boundary conditions but without surface symmetry breaking \cite{bellaiche2000virtual}. 

The profiles of in-plane tilts, $\p_\parallel(z)$, are shown in Fig.~\ref{Fig2}, for a fully saturated 2DEG introduced via explicit oxygen vacancies (red) and the VCA (black) and a pristine interface with no 2DEG (hollow points).
The STO substrate also displays tilts around the out of plane axis, but these are suppressed in the LAO film due to the in-plane expansion imposed by the clamping to the substrate and are therefore not shown here \cite{SM}. 

For a 2DEG induced via the VCA, there is negligible (although nonzero) tilting in the LAO film and in STO near the interface, resulting in a tilt profile virtually identical to the pristine interface.
For a 2DEG introduced via explicit oxygen vacancies, a very large tilt around the [100] axis is induced at the surface which decays into both LAO and STO (despite the competition with the out-of-plane tilt considered in STO).
The presence of an additional distortion when explicit vacancies are considered indicates that there is an additional direct coupling between oxygen vacancies and tilts, separate from the biquadratic coupling between the 2DEG carrier concentrations and tilts,
which occurs indirectly through the polar mode \cite{stengel_tilt,bennett-LAO}.
We emphasise that the lack of significant tilting in the VCA case cannot simply be due to the DFT calculations lying at a saddle-point, since the tilting is small but non-zero. 
This implies that the direct coupling with vacancies does not just break the symmetry to allow for the tilts, but is also sufficiently strong to enforce the tilting with amplitudes orders of magnitude larger than without the coupling. 

\section{Ginzburg-Landau theory}
Motivated by the first-principles calculations in Fig.~\ref{Fig2}, 
we generalize the GL theory developed in Ref.~\cite{bennett-LAO} to account for the additional direct coupling between tilts and oxygen vacancies at the LAO surface \cite{SM}. We write the free energy per unit volume $\F^{\rm GL}$ as
\beq{eq:F_GL_beta}
\F^{\rm GL} = \F^{\rm GL}_{\rm indirect} + \frac{1}{V}\int_{\rm s}\b \s'\p_{\rm s}'\dd{S}
\eep
The indirect terms in $\F^{\rm GL}_{\rm indirect}$ include the energies associated with the defect formation of surface vacancies and associated electrostatic screening \cite{bristowe_electronic_reconstruction}, 
as well as the tilt double well and associated Ginzburg term, and finally an indirect coupling between the tilts and vacancies through the biquadratic coupling between tilts and polar mode \cite{bennett-LAO}.
Both parameters have been renormalized for convenience:
$\p' = \frac{\p}{\p_0}$, $\p_0$ being the bulk equilibrium tilt in LAO, 
and $\s' = \frac{\s}{P_S}$, $P_S$ being the polarization associated with the discontinuity at the interface. 
Interestingly, our symmetry analysis \cite{SM} identifies a new direct coupling term which is bilinear in the surface tilt, $\p_{\rm s}$, and carrier density provided by the oxygen vacancies to the 2DEG, $\s$, with coupling coefficient $\b$, which has units of energy per unit area.
$V=Ad$ is the volume of LAO, where $A$ is the in-plane unit cell surface area, and $d$ is the thickness of the film; while $d$ is a discrete number of unit cells, we treat it as a continuously varying parameter,
since its effect is equivalent to that of an applied electric field (see a more detailed discussion
in Ref.~\cite{bennett-LAO}). Transitions have been predicted and observed for a given thickness
(already close to the transition) under a varying applied perpendicular field \cite{bi2016electro,yang2020nanoscale} and the results of this paper on the character of the transition are directly transferable to that situation.
Being bilinear, the existence of tilting linearly scales the formation energy of surface vacancies, 
and similarly, 
the existence of vacancies produces forces on the surface oxygens which automatically induces tilting.
In this way, unlike the even order coupling terms important for triggered phase transitions 
\cite{mercy2017structurally}, 
the sign of the $\b$ coefficient is unimportant (simply defining the sense of the tilting with respect to vacancies on a certain oxygen sublattice), 
in connection with the field of improper phase transitions \cite{levanyuk1974improper,bousquet2008improper,benedek2011hybrid}.
We note that while a vacancy every $2\times 2$ cells induces tilts around a [100] axis, in agreement with DFT, our group theory analysis suggests that alternative vacancy orderings can induce tilts around [110], which is the expected tilt pattern observed in LAO films \cite{stengel_tilt}, and hence the case considered in the following.

\begin{figure}[t] 
\centering
\includegraphics[width=\columnwidth]{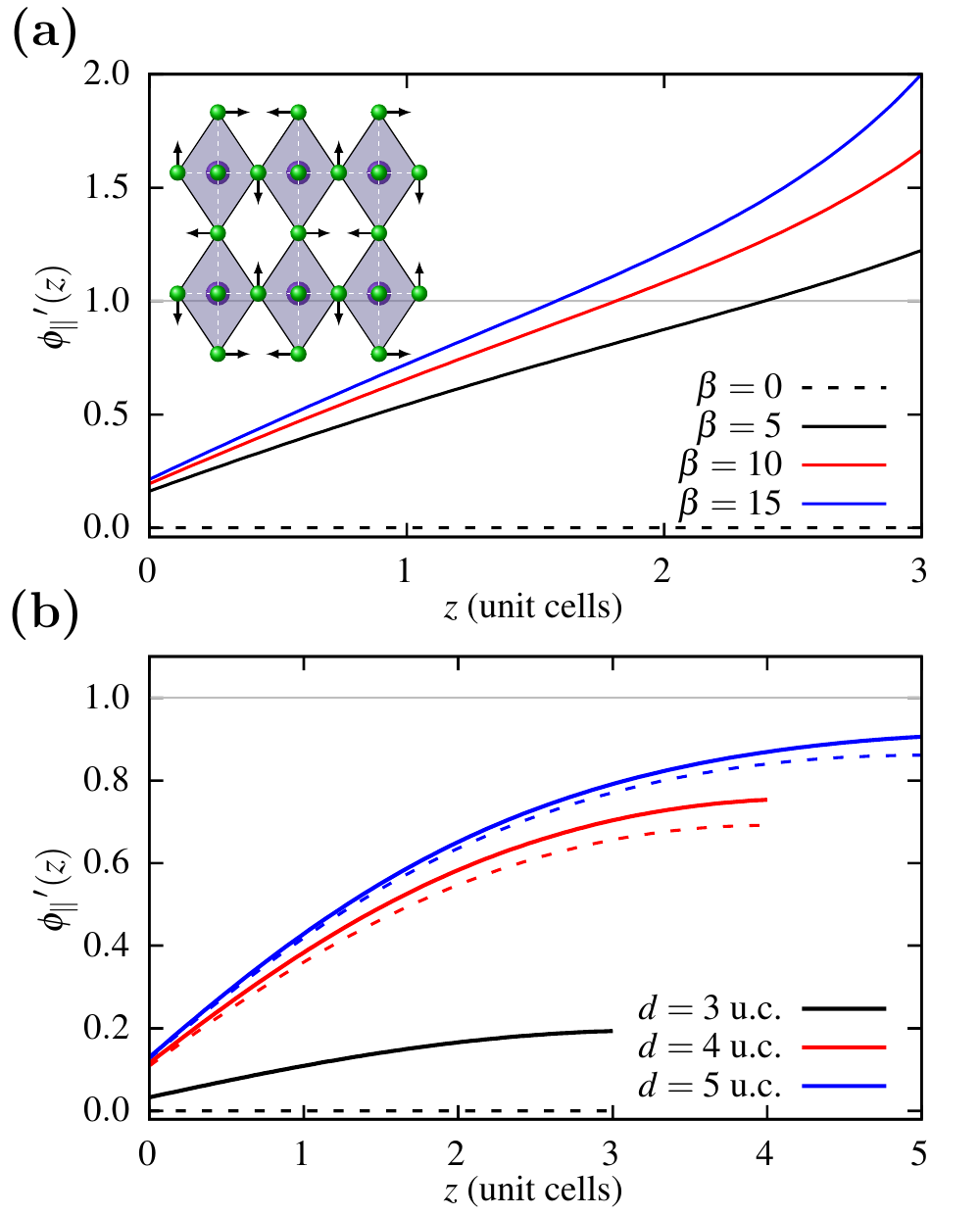}
\caption{
{\bf (a)} Tilt profile for three unit cells of LAO for several values of $\b$,
in units of $\frac{P_S^2}{\ep}$, where $P_S$ is half a polarization quantum of LAO and $\ep$ is the dielectric constant of LAO.
The displacement for the tilt mode around [110], a zone-edge nonpolar mode, is sketched.
{\bf (b)} Tilt profile $\p'(z)$ for LAO films of thickness 3 (black), 4 (red) and 5 (blue) unit cells, 
with ${\b = 0}$ (dashed) and ${\b = \frac{1}{8}}$ (solid). 
In each film the 2DEG is saturated, i.e.~$\s'=1$.
}
\label{Fig3}
\end{figure}

In order to investigate the role of direct coupling between oxygen vacancies and surface tilts on the IMT in LAO/STO, Eq.~\eqref{eq:F_GL_beta} was minimized numerically, following the methodology in Ref.~\cite{bennett-LAO}. 
For a given thickness of LAO and sensible material parameters \cite{SM}, the values of the order parameters $\p'(z)$ and $\s'$ which minimize the free energy were determined.
In Fig.~\ref{Fig3} (a) we show the tilt profile in 3 u.c.~thick LAO for several values of $\b$. 
The magnitude of the the surface tilt which results from the oxygen vacancies is strongly dependent on $\b$, and results in a large tilt profile which decays into the film. For sufficiently strong $\b$, the curvature of the tilt profile changes, and the surface tilt can exceed the bulk equilibrium value, i.e.~$\p'_{\rm s} \equiv \p'(d) > 1$, which is in agreement with the first-principles results shown in Fig.~\eqref{Fig2}.

Fig.~\ref{Fig3} (b) shows the tilt profile in LAO films consisting of 3--5 unit cells, with and without direct coupling between oxygen vacancies and surface tilts.
In the absence of the direct coupling term ($\b = 0$), the 3 u.c.~film is predicted to be untilted, with the tilts appearing from 4 u.c.~and approaching the bulk equilibrium value with film thickness.
Switching on the direct coupling always results in a surface tilt when oxygen vacancies are present, which decays into the film. 
Thus, direct coupling at the surface results in the appearance of tilts in the 3 u.c.~film, without which no tilts are predicted, in line with experimental observations \cite{stengel_tilt}.

\begin{figure}[h] 
\centering
\includegraphics[width=\columnwidth]{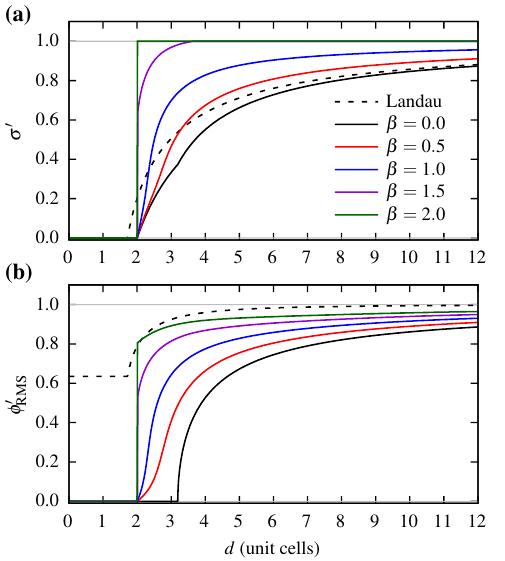}
\caption{
{\bf (a)} Carrier concentration $\s'$, and
{\bf (b)} root mean square of the tilt profile $\p'_{\rm RMS}$, 
as a function of LAO film thickness for several value of $\b$.
The dashed line indicates that the tilts are homogeneous, i.e.~described by a Landau theory, and only the indirect biquadratic coupling between $\s'$ and $\p'$ is considered.
}
\label{Fig4}
\end{figure}

Next, we consider the effect of the direct coupling on the appearance of carriers as a function of film thickness. 
Eq.~\eqref{eq:F_GL_beta} was minimized as a function of film thickness $d$, which was treated as a smoothly varying parameter, for several values of $\b$. The carrier concentration and root mean square tilt $\p'_{\rm RMS} \equiv \sqrt{\left<\p'^2\right>}$ are shown in Figs.~\ref{Fig4} (a) and (b), respectively. Assuming that in the absence of any coupling to tilts, the carriers appear at a critical thickness $\dc = $ 2 u.c.~\cite{sing2009profiling}, the biquadratic coupling to homogeneous tilts (described by a Landau theory) reduces $\dc$ slightly, accompanied by a continuous increase in tilts. Treating the tilts with a GL theory, but with $\b = 0$, the tilts are predicted to appear above 3 u.c., with the carriers appearing at the uncoupled thickness $\dc =$ 2 u.c.. For $\b \neq 0$, the tilts always appear simultaneously at $\dc =$ 2 u.c. when the carriers appear.
The rate at which the carrier concentration increases with thickness is strongly dependent on $\b$.
For sufficiently strong $\b$, both order parameters switch on discontinuously at $\dc$, 
with the $\s'=1$ being fully saturated, 
resembling an avalanche-like phase transition \cite{perez2008multiple,etxebarria2010role}. 

\section{Discussion and conclusions} 
In this letter, we illustrate with first-principles calculations that there is a direct coupling between surface tilts and oxygen vacancies in LAO/STO, 
going beyond the indirect biquadratic coupling of the tilts and the carriers, 
through the polar mode \cite{bennett-LAO}.
We propose that this coupling is described by a bilinear term in the surface tilt and carrier concentration, supported by group theory analysis.
Using a GL theory, we show that the direct coupling always implies the appearance of tilts whenever there are oxygen vacancies. 
This is in agreement with experimental observations \cite{stengel_tilt}, 
as well as our first-principles calculations, where thin films of LAO were predicted to be untilted when interface carriers were artificially introduced using the VCA, but tilted when carriers were introduced from surface oxygen vacancies.

Our results suggest that the direct coupling between oxygen vacancies and tilts on the surface of LAO may play an important role in the IMT at the LAO/STO interface.
Our GL theory suggests that this direct coupling strongly influences the appearance of carriers with film thickness.
The direct coupling between oxygen vacancies and surface tilts may resolve differences between theory, where a continuous appearance of carriers above 3 u.c.~was predicted \cite{bristowe_electronic_reconstruction}, and experiment, where a fully saturated 2DEG is observed at 3 u.c.~and above \cite{ohtomo2004}.

Bilinear coupling terms are typically referred to as pseudo-proper \cite{lines2001principles} since by definition both modes transform with the same irreducible representation, and are typically indistinguishable.
For sufficiently strong coupling, an avalanche-like phase transition to a fully saturated 2DEG occurs.
To our knowledge, this mechanism of an avalanche-like IMT driven by the indirect coupling between surface tilts and oxygen vacancies is unique, adding to the diverse structural mechanisms for IMTs, 
such as the triggerd IMT in nickelates, induced by the cooperate coupling between tilts and a breathing mode \cite{mercy2017structurally}.

We observed a discrepancy between experimental data \cite{stengel_tilt}, where a relatively small surface tilt was measured in 3 unit cells of LAO, and our DFT results, where a large surface tilt was calculated. 
For the same set of material parameters, the tilts from DFT are well-described by a large value of $\b$, Fig.~\ref{Fig3} (a), while the experimental measurements of tilts are better described by a smaller value of $\b$,  Fig.~\ref{Fig3} (b).
This difference in behaviour may be due to the different conditions in ideal DFT calculations and experiment, resulting in very different surface tilts. In Ref.~\cite{SM} we investigate the role of the surface tension on the surface tilt and show that both experimental and DFT results can be described using the same value of $\b$ but changing the energetics at the surface of LAO, keeping all other material parameters fixed.
Despite the differences between the experimental measurements and the more ideal situation in the DFT calculations, the direct coupling between tilts and oxygen vacancies is capable of describing the behaviour of the tilts in both scenarios, as well as a first-order IMT.

The bilinear coupling proposed here may play a role in IMTs beyond polar--nonpolar interfaces. 
A defect ordering described by a non-zone-center mode will couple to non-zone-center phonon modes (see Fig.~\ref{Fig3} (a)), which typically compete with the polar mode.
Thus, for a structural IMT, induced by the suppression/enhancement of a polar mode, this indirect coupling can play a strong role, beyond acting as a source of carriers, in inducing an IMT.

In addition to its role in the IMT at polar--nonpolar interfaces, the direct coupling between soft-mode instabilities and surface defects may prove interesting and useful in other scenarios. 
For example, knowledge of this coupling may make it possible to engineer a regular array of surface defects by tailoring the soft-mode instabilities in a material, through tolerance factor, temperature or strain \cite{tilt_strain}, for example.

The authors thank M.C.~Payne, M.~Stengel and Ph.~Ghosez for helpful discussions.
D.B.~acknowledges funding from the EPSRC Centre for Doctoral Training in Computational Methods for Materials Science under grant number EP/L015552/1, the US Army Research Office (ARO) MURI project under grant No.~W911NF-21-0147 and the National Science Foundation DMREF program under Award No.~DMR-1922172.
P.A.P.~acknowledges funding from the Diputaci\'on Foral de Gipuzkoa through Grant 2020-FELL-000005-01.
Funding from the Spanish MCIN/AEI/ 10.13039/501100011033 is also acknowledged, through grants PID2019-107338RB-C61 and PID2022-139776NB-C65, as well as a Mar\'{\i}a de Maeztu award to Nanogune, Grant CEX2020-001038-M.



%

\clearpage



\section*{Supplementary material}

# SUPPLEMENTARY MATERIAL
# Pseud-proper two-dimensional electron gas formation

Daniel Bennett,[1,2,*] Pablo Aguado-Puente,[3] Emilio Artacho,[1,3,4] and Nicholas C. Bristowe[5]

[1]*Theory of Condensed Matter, Cavendish Laboratory, University of Cambridge, J. J. Thomson Avenue, Cambridge CB3 0HE, UK*
[2]*John A. Paulson School of Engineering and Applied Sciences, Harvard University, Cambridge, Massachusetts 02138, USA*
[3]*CIC Nanogune and DIPC, Tolosa Hiribidea 76, 20018 San Sebastian, Spain*
[4]*Ikerbasque, Basque Foundation for Science, 48011 Bilbao, Spain*
[5]*Centre for Materials Physics, Durham University, South Road, Durham DH1 3LE, UK*
(Dated: September 24, 2024)

## FIRST-PRINCIPLES CALCULATIONS

First-principles calculations were performed using the SIESTA code [1], which employs a basis of localized numerical atomic orbitals (NAOs) [2]. Real space integrations were performed on a uniform real space grid with an equivalent plane wave cutoff of 800 Ry, and a Monkhorst-Pack sampling [3] equivalent to $12 \times 12 \times 12$ in a five-atom perovskite unit cell was used for Brillouin zone integrations. Calculations were carried out within the local density approximation. For geometry optimizations, the in-plane lattice parameter was fixed to the theoretical value for bulk STO ($a = 3.896$ Å), and a threshold of 0.01 eV/Å was used for the forces. All atomic positions were allowed to relax, except for the layer at the bottom of the STO substrate, whose internal coordinates were fixed to the bulk structure in order to mimic the presence of a semi-infinite substrate.

A Hubbard $U$ term [4] was added to the Ti $3d$ orbitals of STO in order to obtain a better approximation to the band gap and avoid underestimating the layer breakdown in LAO with film thickness. A value of $U = 8.5$ eV was found to produce a band gap of 2.95 eV, close to the experimental value of 3.2 eV [5, 6]. Such a large value greatly underestimates the relative permittivity and polarizabilitiy of STO, which is important for the tail of the 2DEG deep in the substrate, where the displacement field decreases and the dielectric constant massively increases [7]. Consideration of this effect goes beyond the scope of this work. Additionally, the tail of the 2DEG is truncated in our calculations due to the finite thickness of the STO slab.

Calculations were performed on vacuum-terminated $(SrTiO_3)_n/(LaAlO_3)_m$ slabs, where $n = 8, 15$ unit cells, and $m = 3$ unit cells. At room temperature, STO has an ideal undistorted perovskite structure. Below 105 K, tilts emerge and STO evolves into a tetragonal phase, facilitated by a rotation of oxygen octahedra in a $a^0 a^0 c^-$ pattern, in Glazer notation [8]. This gives rise to two possible orientations of the tetragonal axis with respect to the interface: the tetragonal axis perpendicular (*Z*-domain) or parallel (*X*-domain) to the interface. Here we focus on the analysis of the *Z*-domain, since the out of plane tilts of the substrate are strongly suppressed in the LAO layers due to the in-plane expansion imposed by the clamping to the substrate. This allows the decoupling of the tilts induced by the presence of surface defects from distortions in the substrate. This domain also better represents the case of $T > 105$ K, when the tilts in STO vanish.

On top of STO, the in-plane lattice parameters of LAO are strongly expanded (+3.3%) with respect to bulk. According to our simulations on strained LAO this results in a suppression of out of plane rotations and a resulting tilting pattern of $a^- a^- c^0$ with an tilt angle of 5.75°, in good agreement with experimental results [9, 10] and previous theoretical studies [11]. At the interface, though, this tendency is disrupted by competition with other distortions, such as polarization due to the unscreened polar discontinuity, tilts from the substrate, and surface defects.

The electrostatic boundary conditions for the LAO film are determined by the presence of free charge at the free surface and interface with STO, coming from (i) an intrinsic breakdown due to the electric field resulting from the polar discontinuity [12] or (ii) the oxygen vacancies at the free surface. The vacancies can be introduced by removing one of the oxygen atoms from the $AlO_2$ surface; removing one oxygen atom from a $2 \times 2$ in-plane supercell is equivalent to imposing a displacement field of $D = 0.5$ electrons per unit cell surface area. In order to distinguish between the effect an oxygen vacancy has on just the electrostatic boundary conditions, and overall, i.e. also on the tilts, we use the virtual crystal approximation (VCA) to include 'virtual atoms' with fractional atomic numbers to gradually modify the electrostatic boundary conditions of the slab [13]. In this case, substituting the oxygen atom at the surface with $O_{1-x}F_x$ can be used to mimic the presence of charged defects at different concentrations.

The tilt profiles for the *Z*-domain are shown in Fig. 2 of the main paper and Fig. S1 here, for rotations in-plane and out-of-plane, respectively. We observe that the explicit vacancies, in the occupation pattern used here of one oxygen vacancy per $2 \times 2$

---

[*] dbennett@seas.harvard.edu

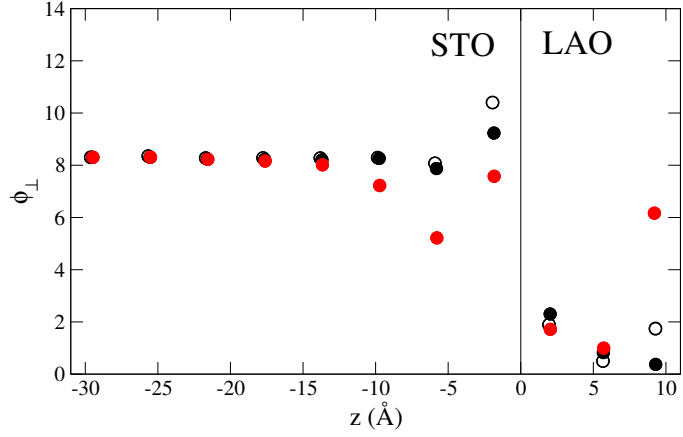

FIG. S1. $\phi_\perp$ tilt profile in LAO/STO from first-principles calculations. The hollow points show results with no 2DEG ($\sigma' = 0$), the filled points show results with a fully saturated 2DEG ($\sigma' = 1$) using VCA (black) and explicit oxygen vacancies at the free surface of LAO (red).

in-plane supercell, produce a tilt of the underlying oxygen octahedra around the [100] axis. In Fig. 2 of the main paper we compare the tilts at the interface with the magnitude of tilts in bulk strained LAO when rotations are constrained to occur around the [100] axis. In Fig. S1 we can see that for the Z-domain calculations the $\phi_\perp$ tilt is approximately constant on the STO side of the interface and then rapidly decays on the LAO side. The reason for this is the tendency of LAO to only display in-plane tilts under these mechanical conditions. Both the uncompensated interface and the VCA calculation show almost identical profiles. The interface with the 2DEG created by explicit oxygen vacancies display a small dip in the oxygen octahedra rotations on the STO side near the interface. We attribute this to the competition with the out of plane tilts induced by the surface defects, whose tail penetrates a small number of unit cells into the STO side.

## GINZBURG-LANDAU THEORY

Here we generalize the Ginzburg-Landau theory developed in Ref. [14] to account for the additional direct coupling between tilts and oxygen vacancies at the LAO surface:

$$\begin{aligned}
\mathcal{F}^{\text{GL}} = & \, 2X_\phi \frac{P_s^2}{\varepsilon} \left( \frac{d_c}{d}\sigma' + \frac{1}{2}\left(1 - \sigma'\right)^2 \right) \\
& + \frac{1}{V}\int_V \left[ \frac{1}{4}\phi'^4 - \frac{1}{2}\left(1 - 2AX_\phi(1-\sigma')^2\right)\phi'^2 - \lambda^2 \phi' \nabla^2 \phi' \right] dV \\
& + \frac{1}{V}\int_s dS \left[ \lambda^2 \left(\hat{n}\cdot \nabla \phi'_s\right) \phi'_s + \frac{1}{2}\delta_s \phi'^2_s + \beta \sigma' \phi'_s \right] \\
& + \frac{1}{V}\int_i dS \left[ \lambda^2 \left(\hat{n}\cdot \nabla \phi'_s\right) \phi'_s + \frac{1}{2}\delta_i \phi'^2 \right]
\end{aligned} \quad (S1)$$

The first line contains the energy associated with a 2DEG of carrier concentration $\sigma' = \frac{\sigma}{P_s}$ [15], where $d_c = \frac{\varepsilon_0 \Delta}{P_s}$ is the critical thickness at which carriers appear, $\Delta$ is the defect formation energy, $\varepsilon$ is the permittivity of LAO and $X_\phi$ represents the stiffness of the tilts, i.e. the curvature about the double well minima. The next line represents the energy of the tilts $\phi' = \frac{\phi}{\phi_0}$ in LAO, where $\phi_0$ is the bulk equilibrium tilt, described by a double well plus a Ginzburg term which describes spatial variations in the out-of-plane direction where $\lambda$ is the correlation length of the tilts. The tilts renormalize the dielectric response of LAO, which affects the appearance of the 2DEG. Thus, the tilts and carriers are indirectly coupled via the biquadratic coupling between the tilts and the polar mode, described by the coefficient $A$, which has units of energy [14, 16]. The last two lines describe the tilts at the free surface (s) and the interface with STO (i), $\phi'_{s/i}$, where $\delta_{s/i}$ are the extrapolation lengths, describing the relative difference in energies between the tilts at the boundaries and in the bulk.

Motivated by the results from first-principles calculations, i.e. Fig. 2 in the main text, we include a term which directly couples the vacancies to the tilt at the surface: the term is linear in the defect concentration and surface tilt, and has coefficient $\beta$.

The bilinear term in Eq. (S1) is motivated by a group theory analysis of the symmetry breaking which occurs when going from the pristine undistorted AlO$_2$-terminated LAO surface to a vacancy ordered and octahedrally tilted surface. Using the ISODISTORT software [17], we find that a periodic array of 1 oxygen vacancy every $2 \times 2$ in-plane unit cells can be described by three occupational modes, transforming as the X$_1$, X$_3$ and M$_5$ irreducible representations (irreps). Importantly, the anti-phase surface octahedral tilting also transforms as the M$_5$ irrep, which implies that the bilinear direct coupling term leaves the free energy invariant. Our invariants analysis also suggests that there is a trilinear coupling between X$_1$, X$_3$ and M$_5$, meaning that reversal of the sense of the M$_5$ tilting necessarily reverses only one of X$_1$ or X$_3$, which we find simply alters the location of the oxygen vacancies on the LAO surface. This also leads to the straightforward realization that the sign of the $\beta$ coefficient does not matter, but only sets the sense of the tilting for a given vacancy sublattice. Note that the same irreps can describe antiphase tilts around either [100] or [110] and alternative surface oxygen vacancy patterns, just with different order parameter directions, and so the same logic for a bilinear coupling term can hold for both tilt patterns.

Following the methodology in Ref. [14], we minimize Eq. (S1) numerically for several values of $\beta$, using $A = 0.407$, $X_\phi^{-1} = 1.365$, in units of $\frac{P_s^2}{\varepsilon}$, obtained from first-principles calculations of bulk LAO. The values of the other parameters used are: $d_c = 2$ u.c., $\lambda = 2$ u.c., $\delta_i = 10$ u.c., and $\delta_s = 0$ u.c.. The resulting order parameters, namely $\phi'(z)$ and $\sigma'$, are shown in Figs. 3 and 4 in the main text for several values of $\beta$ and film thicknesses $d$.

## ROLE OF SURFACE TENSION

The tilt profiles are shown for different values of $\beta$ in Fig. S2, for a different value of $\delta_s$ in each panel.

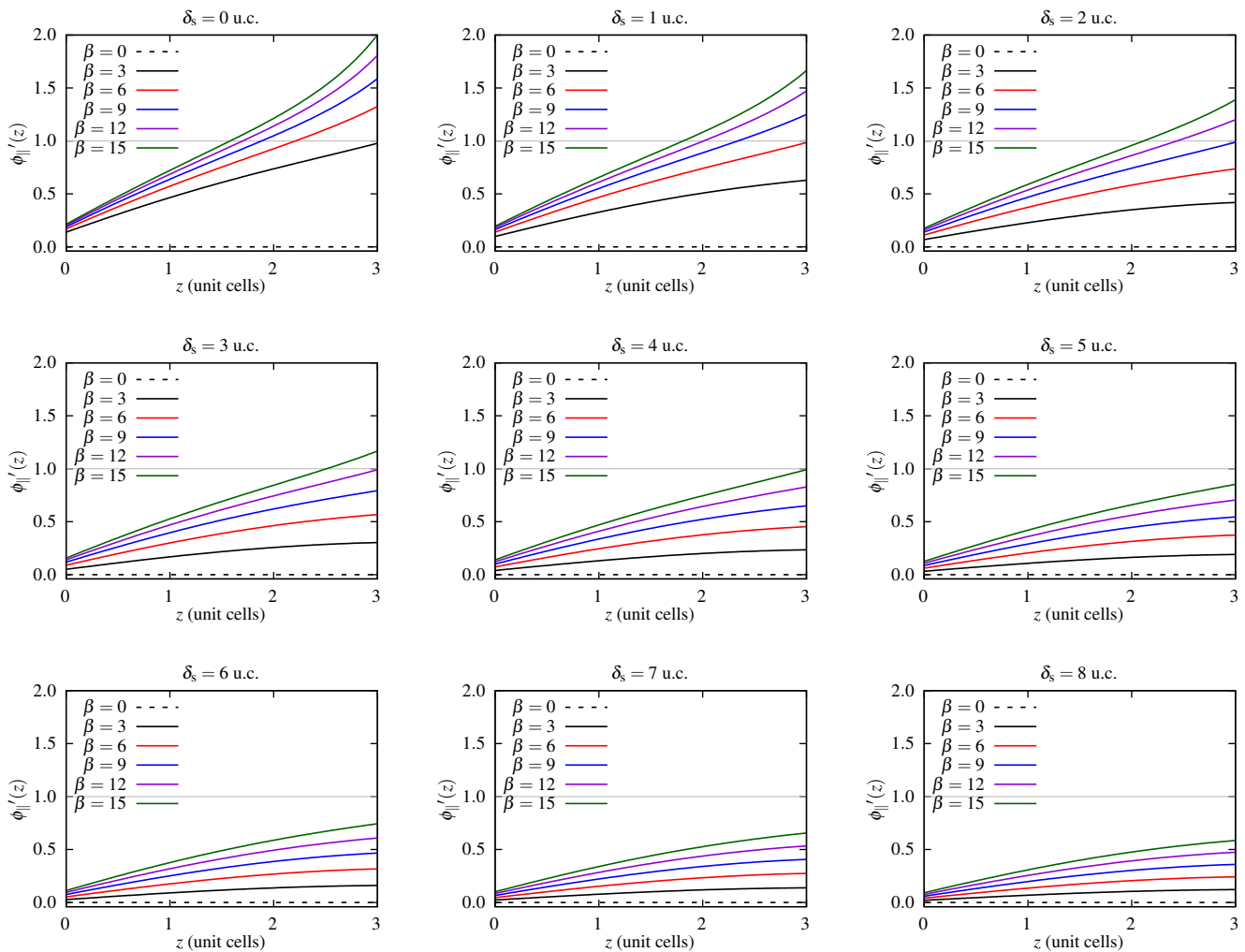

FIG. S2. Tilt profile for three unit cells of LAO as a function of $\beta$ and $\delta_s$, The value of $\delta_s$ in each panel is indicated above.



\end{document}